\documentclass[prl,twocolumn]{revtex4}
\usepackage{epsfig,amsmath}
\usepackage{subfigure}
\usepackage{graphicx}
\usepackage{dcolumn}
\usepackage{stmaryrd}
\usepackage{mathrsfs}
\usepackage{pifont}
\usepackage{amsthm}
\usepackage{amssymb}
\usepackage{bm}
\usepackage{latexsym}
\usepackage{hyperref}
\usepackage{color}
\newcommand{\beq}{\begin{equation}}
	\newcommand{\eeq}{\end{equation}}
\newcommand{\beqa}{\begin{eqnarray}}
	\newcommand{\eeqa}{\end{eqnarray}}

\begin{document}

\author{Zhao-Ming Wang$^{1}$\footnote{wangzhaoming@ouc.edu.cn},Feng-Hua Ren$^{2}$\footnote{renfenghua@qtech.edu.cn}, Marcelo S. Sarandy$^{3}$\footnote{msarandy@id.uff.br}, Mark S. Byrd$^{4}$\footnote{mbyrd@siu.edu}}
\affiliation{$^{1}$ College of Physics and Optoelectronic Engineering, Ocean University of China, Qingdao 266100, China \\
$^{2}$ School of Information and Control Engineering, Qingdao University of Technology, Qingdao 266520, China \\
$^{3}$ Instituto de F\'{\i}sica, Universidade Federal Fluminense, Campus da Praia Vermelha, 24210-346, Niter\'oi, RJ, Brazil \\
$^{4}$ Department of Physics, Southern Illinois University, Carbondale, Illinois 62901-4401, USA}

\title{Nonequilibrium Quantum Thermodynamics in Non-Markovian Adiabatic Speedup}

\begin{abstract} 
Understanding heat transfer between a quantum system and its environment is of undisputed importance if reliable quantum devices are to be constructed.  
Here, we investigate the heat transfer between system and bath in non-Markovian open systems in the process of adiabatic speedup. Using the quantum state 
diffusion equation method, the heat current, energy current, and power are calculated during free evolution and under external control of the system. While the 
heat current increases with increasing system-bath coupling strength and bath temperature, it can be restricted by the non-Markovian nature of the bath. 
Without pulse control, the heat current is nearly equal to the energy current. On the other hand, with pulse control, the energy current turns out to be nearly equal to the power. 
In this scenario, we show that non-Markovianity is a useful tool to drive the system through an approximate adiabatic dynamics, with pulse control acting in the conversion 
between heat current and power throughout the evolution.
	
\end{abstract}

\maketitle

Understanding dissipation phenomena in condensed matter physics is the key to produce reliable nanoscale devices. 
The description of heat exchange, work, and the energy balance due to the interaction between a system and its environment 
is a central topic in nonequilibrium statistical physics \cite{Goold}.  In this context, a seminal contribution is the spin-boson model, 
which provides a clear physical picture for exploring quantum dissipation effects. This model includes an impurity two-level system 
(referred to as a spin) coupled to a thermal reservoir (of bosons), displaying a rich phase diagram near equilibrium \cite{Legget, Hur}.  
The utility of the model has been massively shown by a number of subsequent works, with diversified applications. For example, 
quantum transport has been investigated using a subsystem coupled to two thermal reservoirs in metal-molecule-metal junctions
\cite{Nitzan03}, dielectric-molecule-dielectric systems \cite{Wang07}, electric spin-nuclear spin interfaces \cite{Taylor03}, and metal-superconductor
junctions \cite{Giazotto06}. Moreover, the steady-state heat current in the nonequilibrium spin-boson model has been studied and the effects 
of sampling initial conditions of the thermal baths on the heat current have been found to play an important role \cite{Pablo}. 
Exact dynamics of a class interacting two-qubit systems immersed in separate thermal reservoirs or within a common reservoir has also been studied \cite{Wu2013}. 
Furthermore, the dynamics of the two-spin spin-boson model in the presence of Ohmic and sub-Ohmic baths has been investigated \cite{Deng16}. 
Theoretical investigations about heat transport in a many-body interaction system often use approximation methods, such as a master equation \cite{Wu09}, 
or a Born-Oppenheimer method \cite{Wu11}. Recently, an exactly solvable model was proposed to investigate quantum energy transfer between a nonlinearly 
coupled bosonic bath and a fermionic chain \cite{ZMW20}. 

Nonequilibrium effects play a fundamental role in quantum information processing tasks.
From the point of view of device design, such as in molecular devices, it is necessary to consider the scaling of the energy current with the system size and time in 
order to prevent the devices from disintegrating \cite{Schulze,Pop} due to excess heat exchanged during the operation.  
For example, quantum heat engines require operation within the coherence time of the corresponding platform, which might be very short \cite{Barontini}. 
Thus it is crucial to investigate the heat transfer to and from its surroundings, because the performance of quantum devices depends on optimization control 
protocols aimed at minimizing dissipation \cite{WangX,Dann,Pancotti}. Furthermore, the relaxation  process will cause a certain degree of irreversibility, 
which can be quantified by entropy production, sometimes also called dissipated availability, or excess work \cite{Salamon}.  Energy production limits the 
thermodynamic efficiency of the process \cite{Pietzonka}. Enhancement of the efficiency of heat engines by reinforcement learning approach has recently been 
studied \cite{Sgroi} and successful control of the nonequilibrium quantum process has been realized. 

Thermodynamic processes can be externally driven by quantum control techniques, such as shortcuts to adiabaticity (STA) (for a review, see, e.g., Ref.~\cite{Odelin}).
The STA, or adiabatic speedups, refer to finite-time dynamics with the same final state that would result from infinitely slow, adiabatic driving. 
Possible applications of STA to nonequilibrium quantum thermodynamics have been recently explored \cite{Abah,Moraes}. Moreover, the thermodynamic 
control can be extended to other directions, e.g., quantum annealing \cite{Johnson}. However, a complete analysis of the heat transfer in an open system 
is usually a challenging problem, especially when the bath is non-Markovian \cite{Breuer}, where memory effects of the bath cannot be neglected. 
Different methods have been used to calculate the heat transfer problems in hybrid quantum systems, including hierarchical equations of motion approach \cite{Kato}, 
Redfield theory \cite{Boudjada,Cao2020}, the non-equilibrium Green’s function method \cite{Esposito}, and time-evolving matrix product operators \cite{Popovic}. 
In this letter, we deal with open systems that can be treated using the quantum state diffusion equation method \cite{Diosi98} in the study of nonequilibrium 
thermodynamics of a quantum adiabatic speedup process. In particular, we accelerate the dynamics through cutting off a spin chain and derive a general 
control condition for sine pulse control. Specifically, we obtain the relation between the heat current and the system's reduced density matrix, which 
allows for the use of the heat current as a witness for the quantum adiabatic evolution. Moreover, we will provide a method  {\sl to control the adiabatic speedup} 
via nonequilibrium thermodynamics.
\begin{figure}[]
	\centerline{\includegraphics[width=1.0\columnwidth]{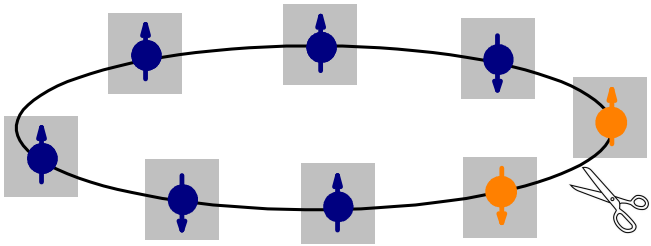}}
	\caption{(Color online) Schematic illustration of cutting a closed spin chain. The chain is immersed in non-Markovian and finite temperature individual heat baths.}
	\label{Fig:1}
\end{figure}

\emph{Model.}---Suppose a small system $H_{s}$ composed of $N$ qubits and immersed in a surrounding environment. Under realistic conditions, 
each qubit will interact with its own heat bath (see Fig.~(\ref{Fig:1})). The Hamiltonian of this open system can be written as
\begin{equation}
H=H_{s}+H_{b}+V_{sb},
\end{equation}
where $H_{s}$ is the system Hamiltonian and $H_{b}=\sum_{j=1}^{N}H_{b}^{j}$ is the sum of $N$-independent bath
Hamiltonians, with $H_{b}^{j}=\sum_{k}\omega _{k}^{j}b_{k}^{j\dag }b_{k}^{j}$ the $j$th bath Hamiltonian ($j=1,2,...,N$), 
$\omega _{k}^{j}$ the boson's frequency of the $k$th mode, and $b_{k}^{j\dag }$ ($b_{k}^{j}$) the bosonic creation (annihilation) operators. 
The interaction Hamiltonian between the system and the baths $V_{sb}$ can be written as
\begin{equation}
V_{sb}=\sum\limits_{j}V_{sb}^{j}=\sum\limits_{j,k}(g_{k}^{j\ast
}L_{j}^{\dag }b_{k}^{j}+g_{k}^{j}L_{j}b_{k}^{j\dag }), 
\end{equation}
where the operator $L_{j}$ describes the coupling between the $j$th qubit in the system and its surrounding bath (defined as the $j$th bath) 
and $g_{k}^{j}$ is the coupling constant between the $j$th qubit and $k$th mode of the $j$th bath.
Assume that all $N$ independent baths are in a thermal equilibrium state and the system's Hamiltonian is in the ground state $\left\vert \mathbf{\psi }_{0}\right\rangle $. 
The density operator for the $j$th bath is $\rho _{j}(0)=e^{-\beta H_{b}^{j}}/Z_{j},$
where $Z_{j}=$Tr$[e^{-\beta H_{b}^{j}}]$ is the partition function and $\beta
=1/(K_{B}T_{j})$, with $T_{j}$ denoting the bath temperature. The initial density operator of the whole system is then taken as a product state
$\rho (0)=\rho _{s}(0)\otimes \rho _{b}(0)$, 
where $\rho _{s}(0)=\left\vert \mathbf{\psi }
_{0}\right\rangle \left\langle \mathbf{\psi }_{0}\right\vert$ and $\rho _{b}(0)=\bigotimes\limits_{j=1}^{N}\rho _{j}(0)$ are the system and bath density
matrices, respectively. The non-Markovian master equation governing the system dynamics can be written as~\cite{WangJPA2021}
\begin{eqnarray}
\frac{\partial }{\partial t}\rho _{s} &=&-i[H_{s},\rho
_{s}]+\sum\nolimits_{j}\{[L_{j},\rho _{s}\overline{O}_{z}^{j\dag
}(t)]-[L_{j}^{\dag },\overline{O}_{z}^{j}(t)\rho _{s}]  \notag \\
&&+[L_{j}^{\dag },\rho _{s}\overline{O}_{w}^{j\dag }(t)]-[L_{j},\overline{O}_{w}^{j}(t)\rho _{s}]\},  \label{eq020}
\end{eqnarray}
where $\overline{O}_{z,(w)}^{j}=\int_{0}^{t}ds\alpha
_{z,(w)}^{j}(t-s)O_{z}^{j}$ and $\alpha _{z,(w)}^{j}(t-s)$ is the
correlation function. The operator $O$ is an \emph{ansatz} and is assumed to be noise-independent here (see for instance Refs.~\cite{Diosi98,YTPRA}).
In Eq.~(\ref{eq020}), we also set $\hbar=1$. 
The spectral density of the bath is needed in order to obtain the correlation function. For the Lorentz-Drude spectrum, the spectral density is
$J_{j}(\omega )=({\Gamma_{j} }/{\pi }) \, [\,{\omega_{j}}/({1+({\omega_{j} }/{\gamma_{j}})^{2}})\,]$~\cite{Wangchemical10,Ritschel,Meier}.  
Here $\Gamma_{j}$ and $\gamma_{j}$ are real
parameters, with $\Gamma_{j}$ representing the strength of the $j$th pair system-bath coupling and $\gamma_{j}$ being the characteristic frequency of the $j$th bath. 
Then, $\gamma_{j}$ controls the correlation time of the bath and decays as $1/\gamma_{j}$. The larger $\gamma_{j}$, 
the smoother the spectral function and the shorter the time the bath takes to relax to equilibrium. 
Thus, the asymptotic limit $\gamma_{j} \rightarrow  \infty$ corresponds to the Markovian regime~\cite{WangJPA2021}.

If we use a Lorentz-Drude spectrum under high temperature or low frequency approximation, closed equations for the operator $\overline{O}_{z,(w)}^{j}$ have been derived 
to numerically calculate the non-Markovian master equation in Eq.~(\ref{eq020})~\cite{WangJPA2021,Fenghua2020}, yielding
\begin{eqnarray}
	\frac{\partial \overline{O}_{z}^{j}}{\partial t} &=&(\frac{%
		\Gamma_{j}T_{j}\gamma_{j}}{2}-\frac{i\Gamma _{j}\gamma_{j}^{2}}{2}%
	)L_{j}-\gamma_{j}\overline{O}_{z}^{j}  \notag \\
	&&-[iH_{s}+\sum\nolimits_{j}(L_{j}^{\dag }\overline{O}_{z}^{j}+L_{j}%
	\overline{O}_{w}^{j}),\overline{O}_{z}^{j}],  \label{eq027}
\end{eqnarray}
\begin{eqnarray}
	\frac{\partial \overline{O}_{w}^{j}}{\partial t}&=&\frac{\Gamma
		_{j}T_{j}\gamma_{j}}{2}L_{j}^{\dag }-\gamma_{j}\overline{O}%
	_{w}^{j}\notag \\
	&&-[iH_{s}+\sum\nolimits_{j}(L_{j}^{\dag }\overline{O}_{z}^{j}+L_{j}%
	\overline{O}_{w}^{j}),\overline{O}_{w}^{j}].  \label{eq028}
\end{eqnarray}
In the Markovian limit $\gamma_{j} \rightarrow  \infty$, the master equation in Eq.~(\ref{eq020}) reduces to the Lindblad equation~\cite{YTPRA,WangJPA2021}, {\it i.e.}
 \begin{eqnarray}
	\frac{\partial }{\partial t}\rho _{s} &=&-i[H_{s},\rho
	_{s}]\notag \\
	&&+\sum\nolimits_{j}\frac{\Gamma_{j}T_{j}}{2}[(2L_{j}\rho _{s}L_{j}^{\dagger}-L_{j}^{\dagger}L_{j}\rho_{s}-\rho_{s}L_{j}^{\dagger}L_{j})\notag \\
	&&+(2L_{j}^{\dag }\rho _{s}L_{j}-L_{j}L_{j}^{\dag }\rho_{s}-\rho_{s}L_{j}L_{j}^{\dag })].	 \label{eq4} 
\end{eqnarray}
\begin{figure}[]
	\centerline{\includegraphics[width=1.0\columnwidth]{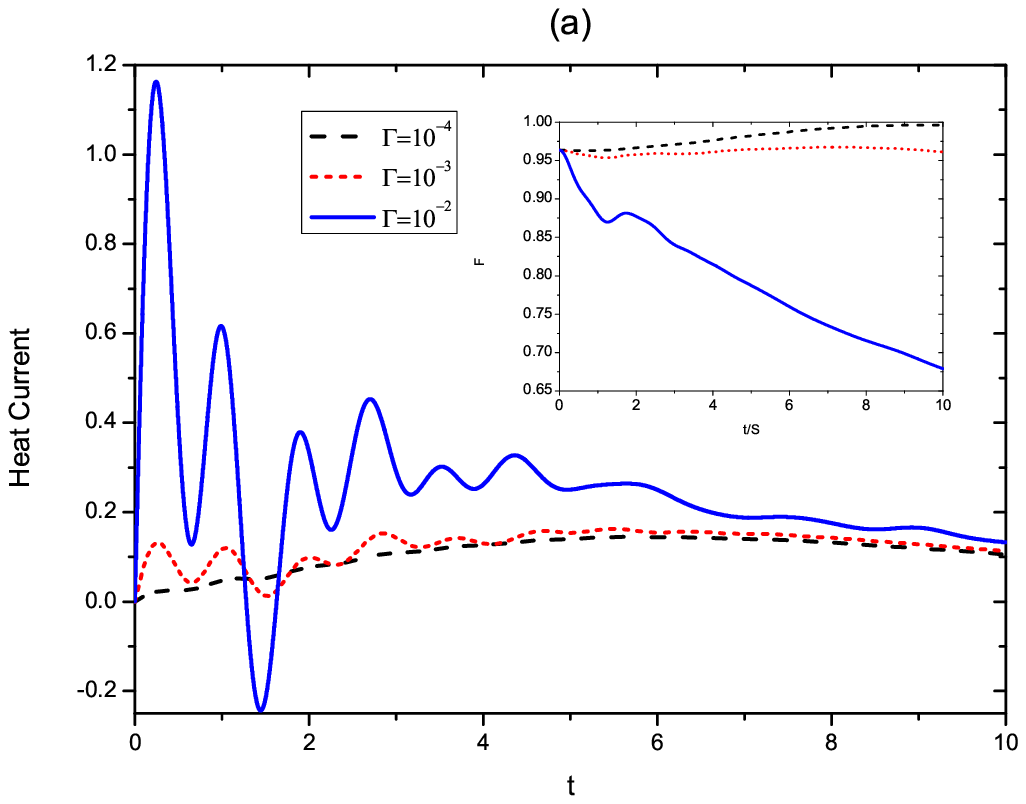}}
	\centerline{\includegraphics[width=1.0\columnwidth]{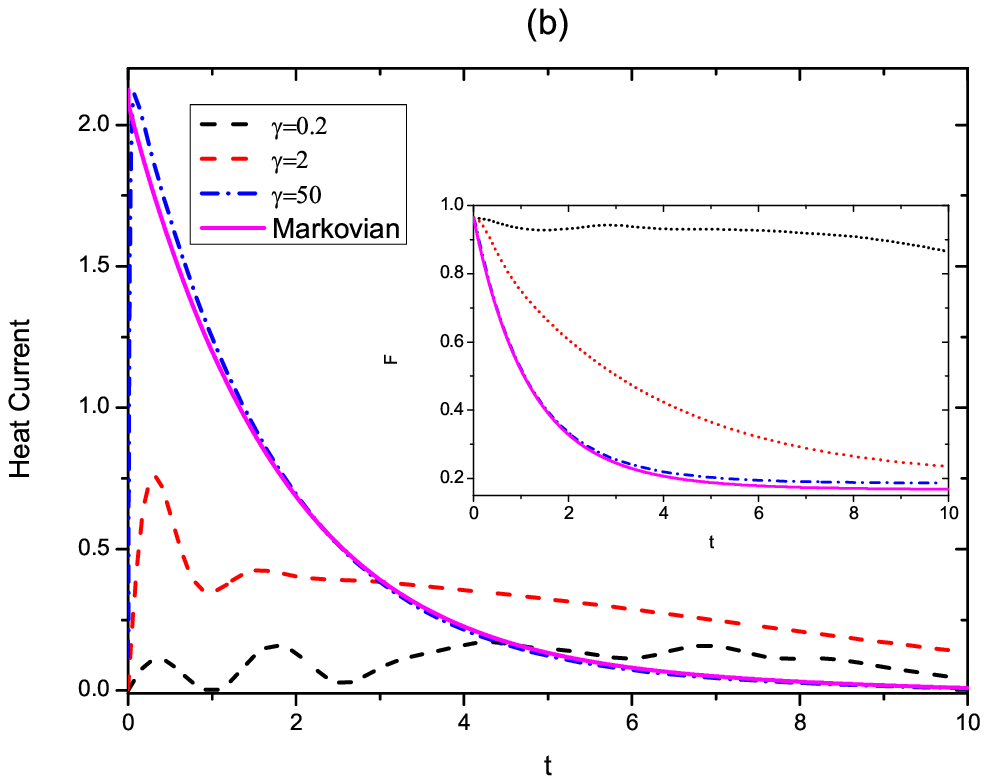}}
	\centerline{\includegraphics[width=1.0\columnwidth]{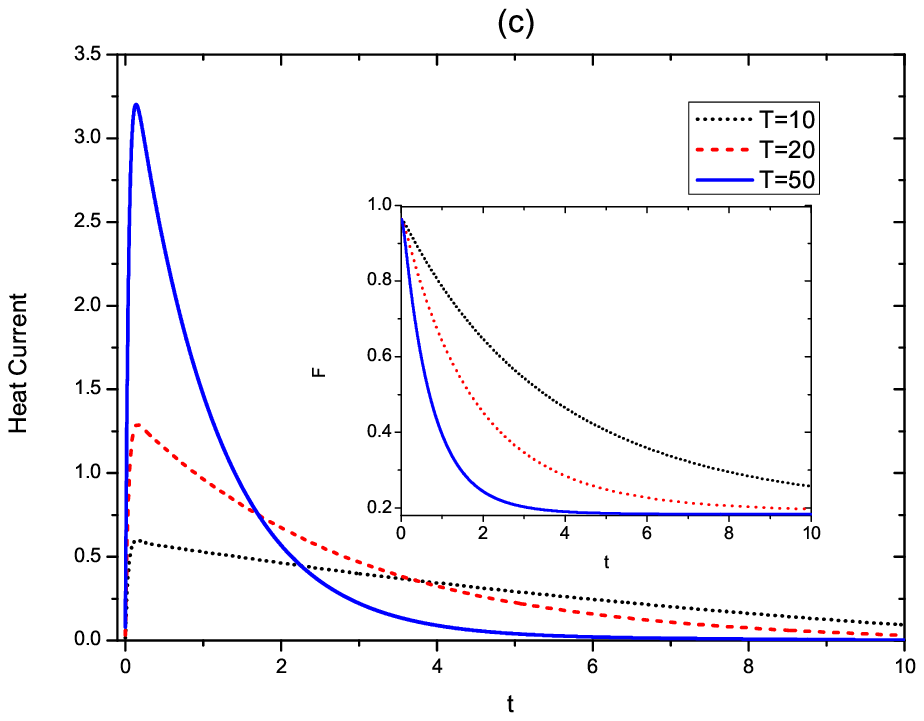}}
	\caption{(Color online) Heat current $J_Q$ as a function of time $t$ for different values of system-bath parameters: (a) Characteristic frequency $\gamma=0.5$ and temperature $T=50$;  
	(b) Strength of the system-bath coupling $\Gamma=0.01$ and temperature $T=30$; (c) Strength of the system-bath coupling $\Gamma=0.01$ and characteristic frequency $\gamma=10$. 
	The total evolution time is $S=10$, the system-bath interaction is described by $L=\sigma_{-}$, and the spin chain has $N=5$ spins. Units are such that $\hbar=1$ and $J=-1$.}
	
	\label{Fig:2}
\end{figure}
For the definition of heat current, we consider two different expressions~\cite{Kato}, one is system heat current (SHC), which is derived through the energy conservation of the system. 
The other is bath heat current (BHC), which is derived through the decreasing rate of the bath energy. The main difference between SHC and BHC is that there is an additional term for the BHC 
when the \textit{j}th and \textit{k}th system-bath interactions are described by non-commuting operators and each system-bath coupling is strong.  Here, in our model, $[V_{sb}^{j},V_{sb}^{k}]=0$ 
and weak-couplings are assumed. Therefore, in this case, SHC is nearly equal to BHC. Hence we use the SHC as the definition of heat current $J_Q$ \cite{Kato}, which reads
\begin{eqnarray}
	J_Q&=&\sum\limits_{j=1}^{N}\frac{dQ_{s}^{j}(t)}{dt}=\frac{d}{dt}\left\langle H_{s}(t)\right\rangle -\frac{dW}{dt}\notag \\
	&& = \left\langle \frac{\partial \rho_{s}(t)}{\partial t}  H_{s}(t)\right\rangle, \hspace{0.5cm}	\label{eq08}
\end{eqnarray}
where $dQ_{s}^{j}(t)/dt=i\left\langle [V_{sb}^{j},H_{s}]\right\rangle$ is the change in the system energy due to the coupling with the $j$th bath. 
Notice also that $dW/dt=\left\langle (\partial H_{s}(t)/\partial t)\right\rangle $ can be defined as the {\it power}, i.e., the time derivative of the work. 
Moreover, $d \left\langle H_{s}(t)\right\rangle/dt$ is the system energy current, {\it i.e.}, the change of the system energy as a function of time.

In what follows, we consider the process of cutting a spin chain, which has been used to study adiabatic speedup \cite{RenPLA,RuiQIP} and non-adiabatic transformations \cite{PyshkinNJP,PyshkinActa}. 
We focus on the heat current and the energy current between the bath and system during the cutting process. The system's Hamiltonian can be written as
\begin{eqnarray}
H_{s}&=&\sum\limits_{i=1,i\neq j}^{N}[J_{i,i+1}(\sigma_{i}^{x
}\sigma_{i+1}^{x}+\sigma_{i}^{y}\sigma_{i+1}^{y})]\notag \\&&+J_{j,j+1}(\sigma_{j}^{x
}\sigma_{j+1}^{x}+\sigma_{j}^{y}\sigma_{j+1}^{y}).
\end{eqnarray}
Suppose the chain is initially closed with the period boundary condition $\sigma_{N+1}^{x,(y,z)}=\sigma_{1}^{x,(y,z)}$. For a uniform XY model, we assume $J_{i,i+1}=J=-1$, $J_{j,j+1}=J\cos \Omega t$, and $\Omega=\pi/(2S)$, 
where $S$ is the total evolution time \cite{RenPLA,RuiQIP}. This cutting has been previously used to study an anomaly in quantum phases induced by borders \cite{JingSR}. In an optical lattice setup, the couplings $J_{j,j+1}$ 
can be tuned individually by focusing additional laser beams perpendicular to the lattice direction \cite{Wang2014pra}. Clearly, when $t=S$, we have $J_{j,j+1}=0$, with the closed chain cut into one open-ended chain (see Fig.~\ref{Fig:1}). 

As an example, let us consider $N=5$ spins and suppose the initial state is the ground state of the Hamiltonian $H_{s}$ in the single excitation subspace at time t=0:  
$\left \vert \psi_{1}(0)\right \rangle=\sqrt{1/5}(\left \vert \textbf{1}\right \rangle+\left \vert\textbf{2}\right \rangle+\left \vert \textbf{3}\right \rangle+\left \vert \textbf{4}\right \rangle+\left \vert \textbf{5}\right \rangle)$~ \cite{RenPLA,RuiQIP}, 
with $\left \vert \textbf{j}\right \rangle$ (j=1,2,...,5) denoting that the state at the $j$th site is a spin up state and all other states are in the spin down state. If the system evolves adiabatically, the final state at time $t=S$ will be $
\left \vert \psi_{1}(S)\right \rangle=(1/2\sqrt{3})\left \vert \textbf{1}\right \rangle+(1/2)\left \vert\textbf{2}\right \rangle+(1/\sqrt{3})\left \vert \textbf{3}\right \rangle+(1/2)\left \vert \textbf{4}\right \rangle+(1/2\sqrt{3})\left \vert \textbf{5}\right \rangle)$.
This is also the instantaneous eigenstate of the time-dependent Hamiltonian $H_{s}(S)$. We observe that there is no degenerate ground state during the cutting process for the chain sizes used as examples in the paper. 
Moreover, we point out that, while exact adiabaticity for closed systems requires an infinitely long time, we have a competition between the evolution slowness and the decoherence time scales in an open system, 
which may lead to an optimal adiabatic fidelity for finite evolution time \cite{Sarandy05,Sarandyprl05,Wangpra2018}. In our work, we will use the fidelity 
$F(t)=\sqrt{\left \langle \psi_{1}(S) \right \vert \rho_{s}(t) \left \vert \psi_{1}(S)\right \rangle}$ with respect to the final state $\left \vert \psi_{1}(S)\right \rangle$ 
expected in the closed system scenario in order to analyze the validity of adiabaticity, with $\rho_{s}(t)$ denoting the system's reduced density matrix in Eq.~(\ref{eq4}).
  
\emph{Heat current without pulse control.}--- 
We now discuss the thermodynamic quantities during the time evolution without pulse control. First we calculate the system-bath heat current as well as the adiabatic fidelity. 
As an example, we take the system-bath interaction as described by $L=\sigma_{-}$, spin chain length $N=5$, and cutting time $S=10$. For these $N$-independent baths, 
the parameters are taken all the same, {\it i.e.}, $\Gamma_{i}=\Gamma$, $\gamma_{i}=\gamma$, $T_{i}=T$ for $i=1,2,...,N$ \cite{Fenghua2020,WangJPA2021}. The heat current $J_Q$ is 
plotted as a function of time $t$ for several different values of $\Gamma$, $\gamma$ and $T$ in Fig.~\ref{Fig:2}(a), \ref{Fig:2}(b) and \ref{Fig:2}(c), respectively. In the inset of 
Fig.~\ref{Fig:2}, we also plot the evolution of the corresponding fidelity. In Fig.~\ref{Fig:2}(a), the maximum of $J_Q$ increases as we increase $\Gamma$, while the fidelity $F(t)$ shows an opposite behavior. 
The environmental parameters are taken to be $\gamma=0.5$ and $T=50$. Therefore, within a non-Markovian setup, stronger couplings  $\Gamma$ are destructive for adiabaticity and exhibit a higher heat current. 
Clearly, the heat transfer oscillation becomes stronger with increasing $\Gamma$ and the heat current sometimes can even reverse from system to bath (see, e.g., $\Gamma=0.01$). 
We also observe that, for large $t$, the heat current decreases independently of the adiabatic fidelity, which strikingly illustrates the distinction between thermodynamic and quantum-mechanical adiabaticity~\cite{Hu:20}.
Fig.~\ref{Fig:2}(b) shows the effects of the characteristic frequency $\gamma$ on the heat current and adiabatic fidelity, with the other bath parameters set to $\Gamma=0.01$ and $T=30$. 
Notice the heat current increases as we increase $\gamma$, with the adiabatic fidelity also showing opposite behavior. 
Then, non-Markovianity helps enhance adiabaticity~\cite{RuiQIP}, with an accompanied smaller heat current. The oscillation of the heat current increases as we decrease $\gamma$, 
demonstrating the non-Markovian nature of the bath. 
In Fig.~\ref{Fig:2}(c), we consider the effects of the temperature on the heat current and adiabatic fidelity, with $\Gamma=0.01$ and $\gamma=10$. 
As expected, a higher temperature destroys the quantumness of the system, exhibiting correlation with an worse adiabatic 
fidelity and a higher maximum heat current. In summary, the departure of the adiabatic behavior always corresponds to a higher heat current, with more heat transferred from the bath to the system. 

\begin{figure}[]
	\centerline{\includegraphics[width=1.0\columnwidth]{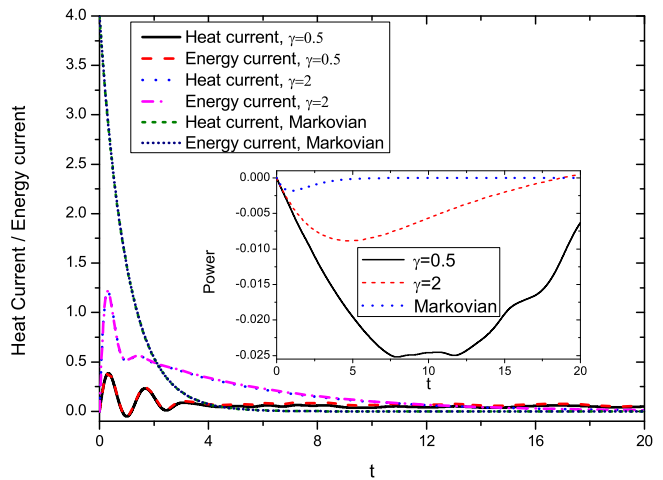}}
	\caption{(Color online) Heat current and energy current as functions of time $t$ for different values of the characteristic frequency $\gamma$. 
	The inset shows the power as a function of time $t$. We set $S=20$, $\Gamma=0.01$, $T=50$, $L=\sigma_{-}$, and $N=5$. Units are such that $\hbar=1$ and $J=-1$. } 
	\label{Fig:3}
\end{figure}

Let us now consider the thermodynamic energy current and the power provided along the cutting process of the chain. In Fig.~\ref{Fig:3} we plot the heat current, energy current, and power as a 
function of time $t$ for different values of the characteristic frequency $\gamma$. The remaining parameters are set to $S=20$, $\Gamma=0.01$, $T=50$, $L=\sigma_{-}$, and $N=5$. We find that the 
power is small compared with the energy current, which indicates that the work on the system can be neglected compared to the change of the system energy during the cutting process. According to Eq.~(\ref{eq08}), 
the heat current is nearly equal to energy current in this case.

\emph{Heat current under pulse control.}---	
 The shortest time possible is desired for quantum information processing tasks so as to avoid as much dissipation and decoherence as possible \cite{Pyshkin16}. Adiabatic speedup has been proposed to achieve an adiabatic evolution in a non-adiabatic regime using control pulses. The system evolution can be controlled by adding an additional Hamiltonian $H_{a}(t)=c(t)M$ to the system Hamiltonian \cite{PierpaoloPRL}, where $M$ is an operator and $c(t)$ is the control function. We let $M=H_{s}(t)$, so the total Hamiltonian can be written as \cite{Fenghua2020,PierpaoloPRL}
	\begin{equation}
		H_{c}(t)=[1+c(t)]H_{s}(t).
	\end{equation}
 Physically, the control function $c(t)$ can be implemented by a sequence of zero-area pulses \cite{Wang20201}. The pulse intensity and period must satisfy certain conditions in order to guarantee an effective adiabatic speedup. Now we consider a general control and derive suitable pulse conditions. Suppose $c(t)$ is a sine function,
 \begin{equation}
 	c(t)=I(b+a \sin\omega t), \label{eq15}
 \end{equation}
 where $a$ and $b$ are the undetermined constant parameters and $I$ is the amplitude of the intensity. By using the one-component Feshbach PQ partitioning technique, the pulse conditions require~\cite{Wang2020}
 \begin{equation}
 	\int_{0}^{\tau}ds \exp[i\int_{0}^{s}c(s')ds']=0. \label{eq16}
 \end{equation}
 Inserting Eq.~(\ref{eq15}) into Eq.~(\ref{eq16}), we obtain
 \begin{equation}
 	\int_{0}^{\tau}ds \exp[i\int_{0}^{s}I(b+a \sin\omega s')ds']=0.
 \end{equation}
  Let $\omega \tau=\pi$, $Ib/\omega=n$ ($n$ is integer), and $Ia/\omega=z$, with $\tau$ being half of the pulse period. Then the pulse condition becomes
 $J_{n}(z)=0$,
 where $J_{n}(z)$ is the $n$th-order Bessel function of the first kind. Clearly, when $b=0, a=1$, the results reduce to our previous zero-area-pulse results, with 
the condition being the zero point of the zero order Bessel function of the first kind~\cite{Wang2020,Chenyf}. 
Notice also that a time-dependent amplitude $I(t)/E_{mn}(t)$ can be used \cite{Fenghua2020} due to the fact that the energy gap between the ground state and the first excited state is time-dependent.

\begin{figure}[]
	\centerline{\includegraphics[width=0.96\columnwidth]{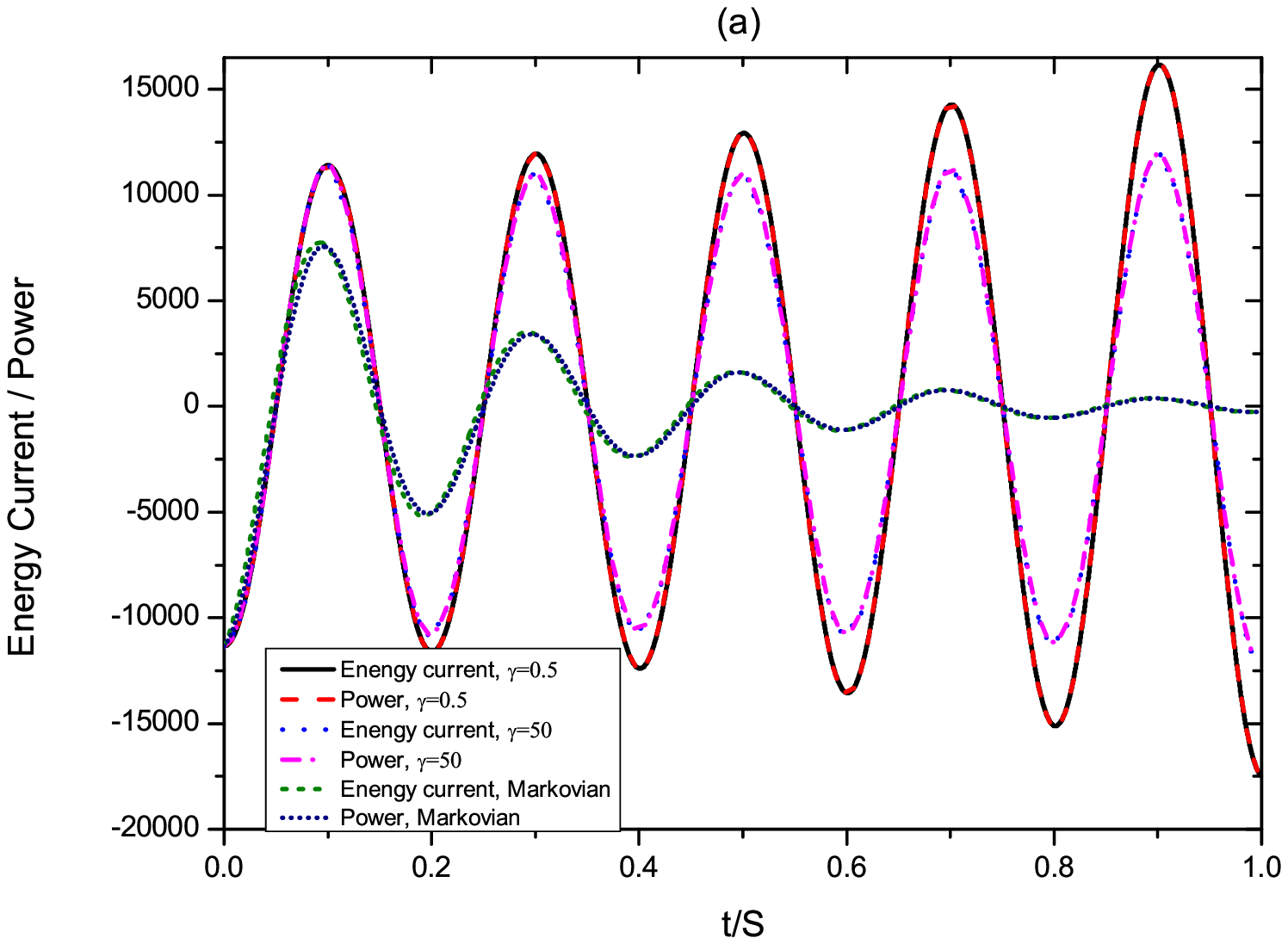}}
	\centerline{\includegraphics[width=0.96\columnwidth]{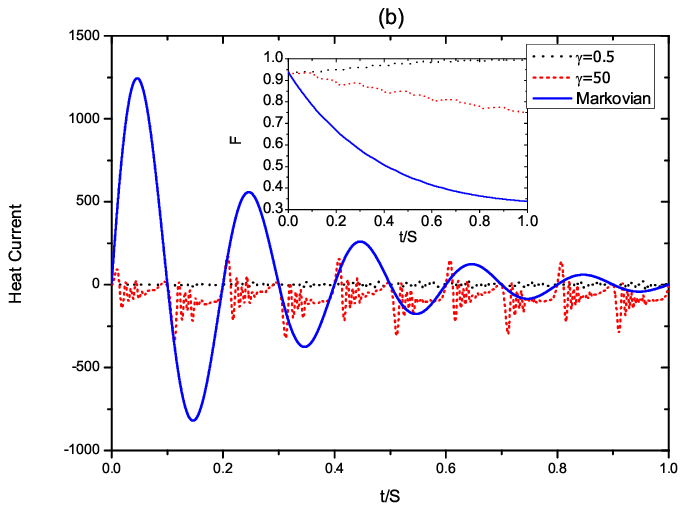}}
	\caption{(Color online) (a) Energy current and power as functions of the rescaled time $t/S$ for different values of the characteristic frequency $\gamma$ under pulse control; 
	(b) Heat current (inset: adiabatic fidelity) as a function of the rescaled time $t/S$ for different values of $\gamma$. The total evolution time is taken as $S=\pi/3$, which provides a non-adiabatic regime. 
	The remaining parameters are set to $T=50$, $I=2.405\times30$, $\tau=\pi/30$, $\Gamma=0.01$, $N=10$, and $L=\sigma_{z}$. Units are such that $\hbar=1$ and $J=-1$.}
	\label{Fig:4}
\end{figure}

Let us consider the heat current, energy current, and power during the process of adiabatic speedup under zero-area sine pulse control. We now take the quantum channel as dephasing $L=\sigma_{z}$ and the other parameters as 
$N=10$, $\Gamma=0.01$, and $T=50$. Moreover, the time evolution is set to $S=\pi/3$, which provides a non-adiabatic regime. The control pulses satisfy $J_{n}(z)=0$, with $I=2.405\times30$ and $\tau=\pi/30$. 
The pulse function is taken to be $c(t)=I\sin(\pi t/\tau)/E_{21}(t)$, with the time-dependent energy gap $E_{21}(t)=E_{2}(t)-E_{1}(t)$. In Fig.~\ref{Fig:4}(a)~and~(b), we plot the energy current, power, and heat current 
as a function of the rescaled time $t/S$ during the adiabatic speedup for different values of the characteristic frequency $\gamma$. In the inset of Fig.~\ref{Fig:4}(b) we plot the fidelity versus time $t/S$. 
Fig.~\ref{Fig:4}~(a) shows that the energy current is nearly equal to the power for small $\gamma$, and they begin to deviate with the increase of $\gamma$. Both the energy current and power oscillate with time 
due to a sequence of periodic driving pulses. In the Markovian case, the current and power become nearly zero at the end of the evolution while for the non-Markovian case ($\gamma=0.5$), the amplitude becomes 
larger and larger. In Fig.~\ref{Fig:4}(b), we plot the corresponding heat current. Clearly the heat current becomes larger as we increase $\gamma$. The same thing happens with the free evolution case, with a 
smaller heat current corresponding to a higher adiabatic fidelity. Thus, the decrease of the heat current provides a {\it signature} of the system adiabaticity. Notice also that non-Markovianity does not only help enhance the 
effects of the pulse control but also constrain the heat current. In this case, the change in the system energy is nearly equal to the work on the system. This is different from the case without control, where the change in the 
system energy is nearly equal to the heat change. 

\emph{Conclusions.}--- 
We have shown how to describe the quantum dissipative dynamics and heat transfer in non-Markovian finite temperature heat baths using the quantum state diffusion approach. By using spin chain cutting method, 
we have been able to describe the adiabatic speedup under the control of external pulses. We have found that the degree of non-Markovianity constrains the heat current. The results show that the energy 
current is nearly equal to heat current without control, while it can be made nearly equal to the power with pulse control. For both cases, a smaller heat current always corresponds to a higher adiabatic fidelity. 
Hence, non-Markovianity can be used to boost adiabaticity, with pulse control acting in the conversion between heat current and power throughout the evolution. 
These tools potentially provide useful applications for the control of energy and information transfer in quantum device engineering.

We would like to thank E. Ya. Sherman for his original idea for cutting the spin chain and recent helpful discussions. This work is supported by Natural Science Foundation of China (Grant No.
11475160), Shandong Provincial Natural Science Foundation, China (Grant No. ZR2014AM023). 
M.S.S. is supported by Conselho Nacional de Desenvolvimento  Científico e Tecnológico  (CNPq) (307854/2020-5).  
M.S.S. also acknowledges financial support in part by Coordena\c{c}\~ao de Aperfei\c{c}oamento de Pessoal de N\'{\i}vel Superior  - Brasil (CAPES) 
(Finance Code 001) and by the Brazilian National Institute for Science and Technology of Quantum Information [CNPq INCT-IQ (465469/2014-0)]. 
M.S.B. is supported by the NSF, MPS under award number PHYS-1820870.

\end{document}